\documentclass{svjour2}                    
\smartqed  
\usepackage{graphicx}
\usepackage{enumerate}
\begin{document}

\title{Growth of scale-free networks under heterogeneous control}

\author{Linjun Li         \and
        Xiangwen Wang 
}

\institute{Linjun Li \and
           Xiangwen Wang\at
              Department of Physics, Virginia Tech, Blacksburg, VA 24061-0435 USA \\
              \email{linjunli@vt.edu}           
}

\date{\today}

\maketitle

\begin{abstract}
Real-life networks often encounter vertex dysfunctions, which are usually followed by recoveries after appropriate maintenances. In this paper we present our research on a model of scale-free networks whose vertices are regularly removed and put back. Both the frequency and length of time of the disappearance of each vertex depend on the degree of the vertex, creating a heterogeneous control over the network. Our simulation results show very interesting growth pattern of this kind of networks. We also find that the scale-free property of the degree distribution is maintained in the proposed heterogeneously controlled networks. However, the overall growth rate of the networks in our model can be remarkably reduced if the inactive periods of the vertices are kept long.

\end{abstract}


\section{Introduction}

In the World Wide Web (WWW), the temporary shutdowns of major websites due to maintenance always bring people great inconveniences. The hyperlink pattern may change during the shutdowns of the important websites, and the path through which people fetch certain information also changes. New websites joining the WWW with preferential attachments may not connect to the websites that are under maintenance, and the growth pattern of the WWW can be accordingly altered. Similar situations also happen in transportation systems, warehouse systems, supercomputing systems, power grids, etc. What is in common is that, if the vertices of the networks are repeatedly deactivated due to maintenance or some other issues, both the growth pattern of the networks and dynamics on the networks become different.

Former research of the growth of networks with interventions majorly focused on extreme cases, in which changes of edges/vertices are made permanently. Typical models related to the intervention of edges include the {\it internal edges and rewiring model} 
\cite{AB} and the {\it internal edges and edge removal model} 
\cite{DM}. Typical models related to the intervention of vertices include the {\it random vertex removal model} and {\it preferential vertex removal model}
\cite{AJB,MGN,BMRS}.
Recently, research progress of network intervention is mainly made in the areas of network controllability \cite{LH,KM,NV,SM} and network control strategies \cite{PPR,CKM}.

Compared to the previous models of controlled networks, in this paper we present our research of networks that are controlled by standardized regulations.
The networks of our model show two important characteristics: first, the vertices in our networks are repeatedly deactivated and reactivated, instead of being removed permanently after the deactivations; second, the control of each vertex is heterogeneous, and dedicated to the specific vertices according to the standardized regulations. The first characteristic guarantees a revival after each network failure, while the second characteristic can drastically complicate the growth behavior of the networks in question.

\section{The Model}

Our model is based on Barab\'asi-Albert model (BA model) which generates scale-free networks with preferential attachments \cite{BAmodel}. To implement the repeated deactivation and reactivation of the existing nodes of the networks, we introduce two important parameters: $\alpha$, denoting the maximum length of time a node can stay active; and $\beta$, denoting the maximum length of time a node can stay inactive. To realize the heterogeneity of network control, we rank the existing nodes in a real-time fashion, and assign the following length of time $t_1$ ($t_2$) for active (inactive) status if they are at the end of their current inactive (active) status. In addition, we assume that if the $t_2$ period for a vertex is finished, this vertex and its attachments (which are built before deactivation) are simultaneously put back, enabling newly created vertices attaching to it according to its formerly established degree. 

\begin{figure} [h]
\includegraphics[width=0.8\columnwidth]{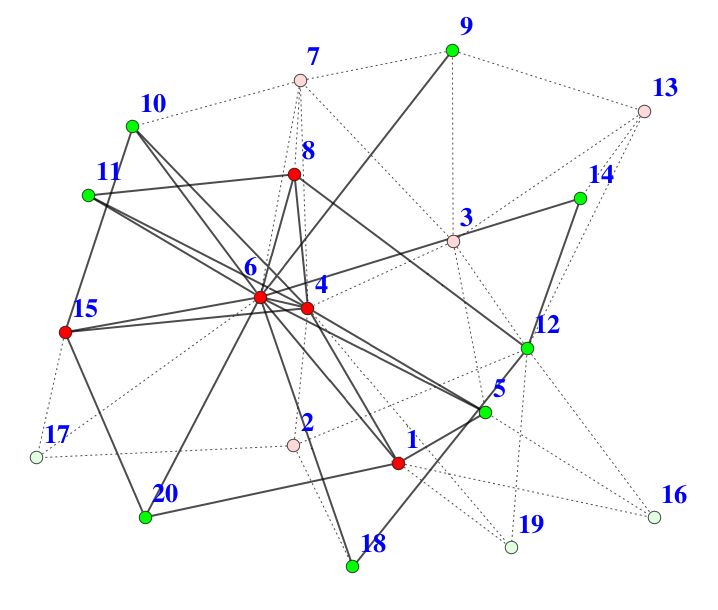}
\caption{\label{fig1} (Color online) Schematic picture of a modeled network at time $t=17$, starting with three disconnected vertices ($m_0=3$). Each new vertex comes with 3 new edges ($m=3$), with the other ends connected to 3 existing verticies. The vertices in the network are divided into two tiers ($n=2$). Vertices colored with dark red (dark green) are active and categorized into tier one (two). Meanwhile vertices colored with light red (light green) are inactive and categorized into tier one (two).  The edges that are attached to a vertex will be removed (shown as dotted lines) if the vertex has been deactivated, and will be brought back once it is reactivated. 
}
\end{figure}

The evolution of our network is illustrated in Fig.\ref{fig1}, and the algorithm for the evolution can be described as follows:
\\1, start with $m_0$ disconnected vertices, then assign $t_1=\alpha$ for each vertex;
\\2, in each time step, check the length of time each vertex has been sitting in its current status:
\begin{enumerate}[i]
\item if its current status is active and this status expires, remove the vertex and the edges that are attached to it;
\item if its current status is inactive and this status expires, put back the vertex and the edges that are formerly attached to it;
\end{enumerate}
3, once the status is changed, we assign $t_1$ ($t_2$) to the vertex according to the following rule:
\begin{enumerate}[i]
\item rank the given vertex:  if $k\in ({k_{max}}^{\frac{n-i}{n}}, {k_{max}} ^{\frac{n-i+1}{n}} ]$ , we put it into tier No.$i$ (, $i=1,2,.. n$, $k_{max}$ is the largest degree of the current network);
\item if the new status of the vertex is active (inactive), assign $t_1=\alpha/i$ ($t_2=\beta/i$) for the vertex to stay active (inactive);
\end{enumerate}
4, create a new vertex, and attach it to $m$ existing vertices, with a probability of $P_s=\frac{k_s}{\sum_{j}k_j}$ for the attachment to the $s^{th}$ vertex;
\\5, continue to the next time step.

\section{Simulations and Results}

To study the growth behavior of the proposed network systems, we first investigate the quantities of total number of edges $L'(t)$ and average geodesic length $l$. Then, we study the overall trend of the growth by calculating normalized growth rate $\bar{k}$.  At last, we show our data for the degree distribution $P(k)$ of the simulated networks.

\subsection{Total Number of Edges and Average Geodesic Length}

From the network evolution algorithm introduced in the last section, we understand that every vertex is deactivated and reactivated repeatedly in our model; we also understand that the edges connected to each vertex are removed and reestablished together with the vertex. As a result, the total number of edges $L'(t)$ is not expected to grow linearly with time $t$ as is described by 
      \begin{equation}
      \label{eq:eq1}
      L(t)=mt+c_0
      \end{equation}
 in the BA model \cite{Rev}, where $c_0$ is the number of edges this network start with ($c_0=0$ in our algorithm); and $m$ is the number of edges each new vertex brings to the network. 

While $L'(t)$ gives us information about the overall connectivity of the network, the average geodesic length $l(t)$ gives us information about the average distance between two randomly chosen vertices in the network \cite{Newman}. In our research, we use Dijkstra's algorithm to compute for the geodesic length $d(v,w)$ between a given vertex pair $w$ and $v$ \cite{DA}. If vertices $w$ and $v$ are not connected, we set $1/d(v,w)=0$. Thus, we implement the computation of $l(t)$ as shown in Eq.\ref{eq:eq2}.
      \begin{equation}
      \label{eq:eq2}
      l(t)\equiv 1/{\left \langle \frac{1}{d(v,w)} \right \rangle}\equiv 1/\left [\frac{1}{N(N-1)}\sum_{v=1}^{N}\sum_{w=1,w\neq v}^{N}d(v,w) \right ]
      \end{equation}

\begin{figure} [h]
\includegraphics[width=1\columnwidth]{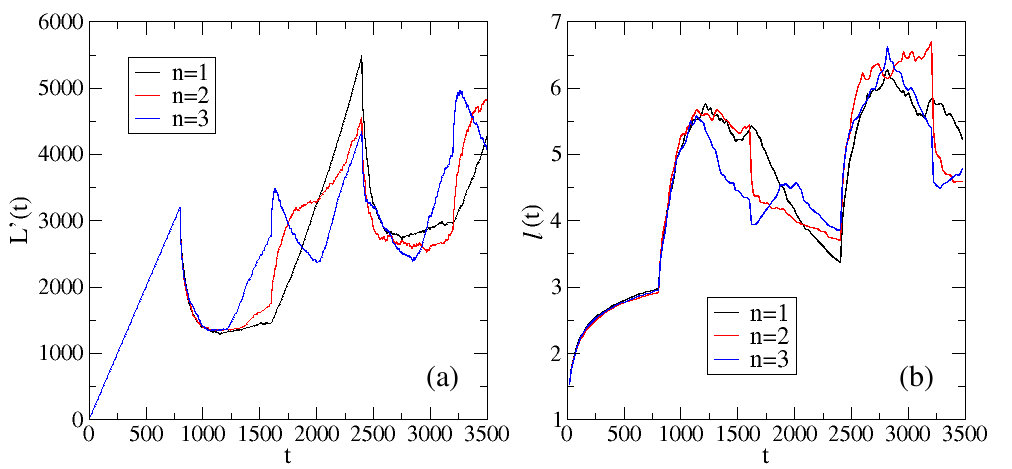}
\caption{\label{fig2} (Color online) Growth of heterogeneously controlled networks with $m_0=m=4$, $\alpha=\beta=$ 800, and $n$ =1, 2 and 3. (a) shows the data for total number of edges $L'(t)$; (b) shows the data for average geodesic length $l(t)$. From both (a) and (b), we find the periodicity of the network growth may change drastically with the change of the total number of tiers. When $n=1$, there are only two troughs (peaks) in the plot of $L'(t)$ ($l(t)$). When $n$ is increased to 3, transition of periodicity is made and there are three troughs (peaks) in the plot of $L'(t)$ ($l(t)$).}
\end{figure}

We use the two quantities introduced above to first study a simple case of the proposed network model, in which $\alpha=\beta$. In Fig. \ref{fig2}, we show the $L'(t)$ and $l(t)$ of typical runs of network evolutions with $\alpha=\beta=$ 800, and total number of tiers n=1, 2, and 3. 

We can see from Fig. \ref{fig2}(a), that our networks automatically revive after each major breakdowns. If we increase the value of n, the periodicity of $L'(t)$ becomes more and more complicated. For example, at the time $t=1600$, the $L'(t)$ for n=1 network starts an abrupt increase, which is a combined result of adding new vertices (with new edges) and reestablishing the formerly deactivated vertices (with old edges). However, if we set n=2, a minor increase starts at $t\approx1360$, indicating some vertices in tier No.2 with $t_2=400$ are starting to reappear at this point, bringing back edges connected to it back to the network. Further, if we set n=3, a large increase of $L'(t)$ starts at even an earlier time, around $t\approx 1200$, and this is believed to be majorly caused by the reactivation of vertices from both tier No.2 and tier No.3. The argument for why this large increase of $L'(t)$ for n=3 network does not happen earlier than t=1200  is that: the vertices in tier No.3 may have been reactivated before t=1200, yet their contributions to increase $L'(t)$ are very small since they all hold small degrees. We can use the same reasoning to explain why there is only a minor increase at $t\approx1360$ for the n=2 network.

In Fig. \ref{fig2}(b), we show the corresponding geodesic length for the evolving networks. After a short period ($t=\alpha$) of logarithmic growth, $l(t)$ starts to increase greatly due to the vertex deactivations, and it starts to decrease after another short period.
A comparison between Fig. \ref{fig2}(a) and Fig. \ref{fig2}(b) shows, each peak of the $L'(t)$ curve can be mapped to a trough of the $l(t)$ curve for the same network, and vice versa. 

In real-world situations, we usually don't expect the vertices go into inactive status too often. Once a vertex is deactivated, we usually don't want to keep it inactive for long period of time. Thus in general, we assume $\alpha>\beta$.
\begin{figure} [h]
\includegraphics[width=1\columnwidth]{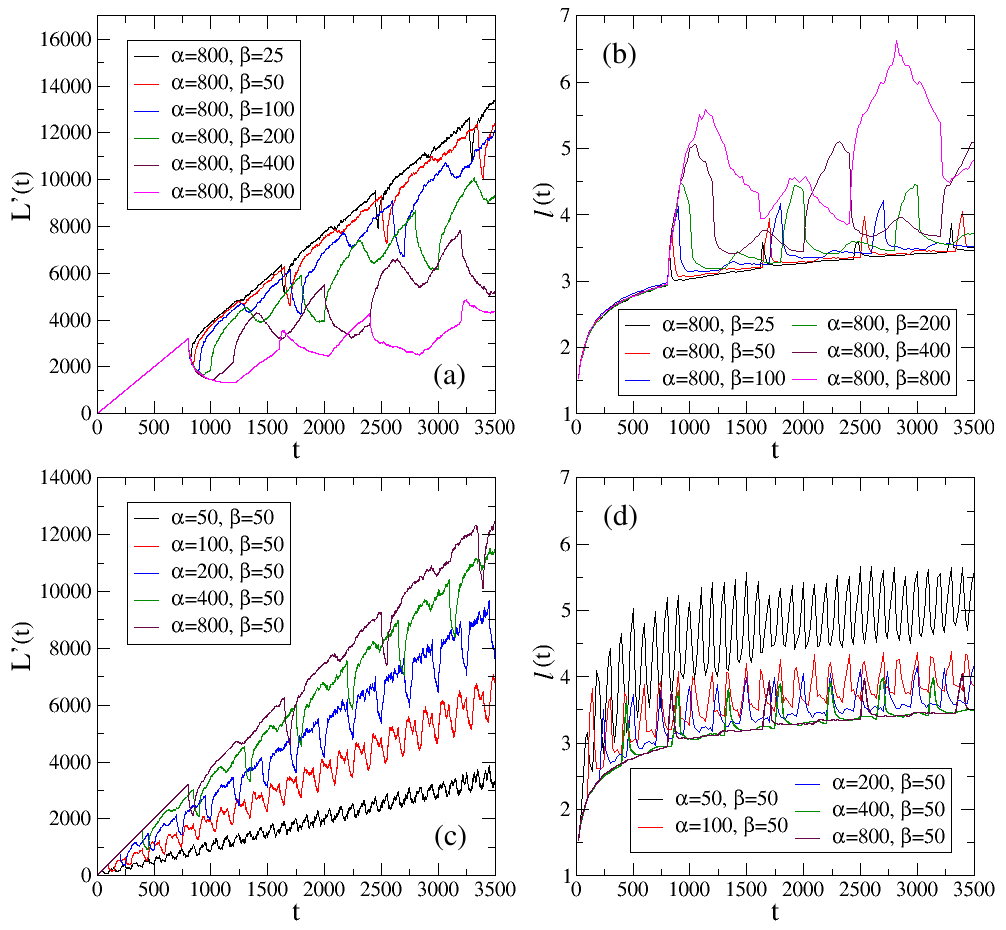}
\caption{\label{fig2_2} (Color online) Growth of heterogeneously controlled networks with $m_0=m=4$, and varied combinations of $\alpha$ and $\beta$. The total number of tiers $n$ =3. (a) and (c) show the growth of $L'(t)$, which is the total number of edges; (b) and (d) show the growth of $l(t)$, which is the average geodesic length. 
}
\end{figure}
In Fig. \ref{fig2_2}, we show how the $L'(t)$ and $l(t)$ evolve when the networks hold different combinations of $\alpha$ and $\beta$ ($\alpha>\beta$). In Fig. \ref{fig2_2}(a) and Fig. \ref{fig2_2}(b), we fix $\alpha$ and increase $\beta$ exponentially. In Fig. \ref{fig2_2}(c) and Fig. \ref{fig2_2}(d), we fix $\beta$ and increase $\alpha$ exponentially. If we define the normalized growth rate as:
      \begin{equation}
      \label{eq:eq3}
 \bar{k}(\alpha,\beta,t)\equiv \frac{k(\alpha,\beta,t)}{k_0}
      \end{equation}
(where $k$ is the slope of the fitted straight line of the data of $L'(t)$, or the overall growth rate of the controlled network; and $k_0$ is the slope of $L(t)$ for the corresponding BA model, or the overall growth rate of the corresponding BA model. In fact, the ${k_0}$ here is equal to $m$.) We can infer from Fig. \ref{fig2_2}(a) and Fig. \ref{fig2_2}(c) that, an increasing ratio of $\alpha/\beta$ leads $\bar{k}$ to approach 1. Also, as shown in Fig. \ref{fig2_2}(b) and Fig. \ref{fig2_2}(d), the average geodesic length $l(t)$ approaches a logarithmic growth pattern when the value of $\alpha/\beta$ is increased. (The logarithmic growth of average geodesic length is a typical characteristic of BA model \cite{Rev}.) Briefly, both of the two findings coming out of Fig. \ref{fig2_2} indicate that the growth behavior of BA model will be recovered if we take the limit of $\alpha/\beta \to \infty$. Otherwise, the growth pattern of $L'(t)$ and $l(t)$  can be different, and the overall growth rate can be remarkably reduced.

\subsection{Normalized Growth Rate}
To confirm the claim made at the end of last section, we systematically simulate for the normalized growth rate $\bar{k}$. 
\begin{figure} [h]
\includegraphics[width=1.0\columnwidth]{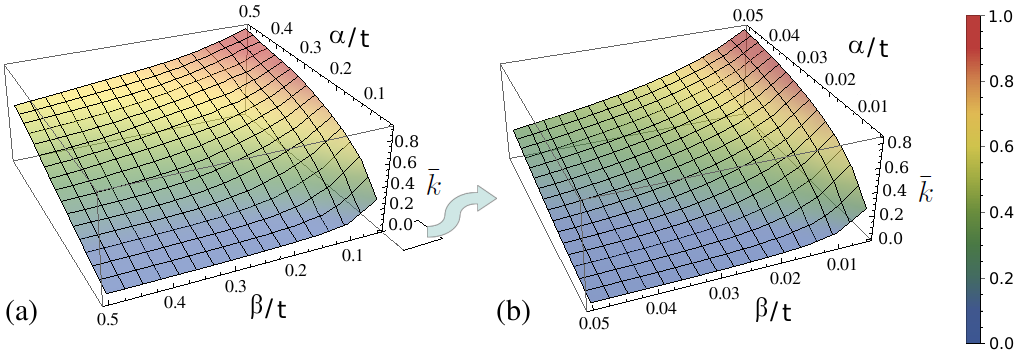}
\caption{\label{fig3} (Color online) Each data point on the plots is an average of 1000 realizations of network generations. The networks are generated with $m_0=m=4$, $n=3$, and the figures are plotted at $t=4000$.(a) $\bar{k}$ data for networks with $\alpha ,\beta$=200, 400, ..., 2000. (b) $\bar{k}$ data for networks with $\alpha ,\beta$=20, 40, ..., 200. Other points on the surfaces are obtained by two-dimensional spline interpolation. 
}
\end{figure}
Fig. \ref{fig3} maps out the relation between $\bar k$ and different combinations of $\alpha$ and $\beta$, in different scales. Each data point on the plots of Fig. \ref{fig3} is an average of 1000 realizations of network generations. It is obvious from both Fig. \ref{fig3}(a) and Fig. \ref{fig3}(b) that $\bar k$ values are relatively small when $\alpha\leq\beta$. For example, the maximum value of $\bar k$ for the $\alpha\leq\beta$ part of Fig. \ref{fig3}(a) happens at $\alpha/t=\beta/t=0.5$, which is equal to 0.54631; while the maximum value of $\bar k$ for the $\alpha\leq\beta$ part of Fig. \ref{fig3}(b) happens at $\alpha/t=\beta/t=0.05$, which is equal to 0.66586. On the contrary, the $\bar k$ can be relatively large or even close to 1 in the region where $\alpha>\beta$. The maximum value of $\bar k$ for the $\alpha>\beta$ part of Fig. \ref{fig3}(a) is found at $\alpha=0.5$ and $\beta=0.05$, which equals 0.90763; while the maximum value of $\bar k$ for the $\alpha>\beta$ part of Fig. \ref{fig3}(b) is found at $\alpha=0.05$ and $\beta=0.005$, which equals 0.83634. From another perspective, it is not hard to find from both of the two figures that a point with larger $\alpha/\beta$ values has larger $\bar k$ values (, which are represented by warmer colors). And indeed we get the largest $\bar k$ value when $\alpha/\beta$ takes the largest values in our simulations. This is a quantitative confirmation that our proposed model may behave more and more like BA model if we continue to increase the ratio of $\alpha$ and $\beta$.


\subsection{Degree Distribution}
In the last two sections, we find that the evolution of $L'(t)$s of the proposed networks are always following a pattern of linear increase with periodic ups and downs, thus we use linear least squares fitting technique to approximate for the data of $L'(t)$s. Then, we defined normalized growth rate $\bar k$ as a ratio of the slope of the fitted line of the $L'(t)$ of our network and the slope of the $L(t)$ of the corresponding BA network. 

In addition, it is not hard to prove for our model that when $t\gg \alpha,\beta$, the $\bar k$ in Eq. \ref{eq:eq3} can be approximated by a function of only $\alpha$ and $\beta$. Thus, we can use $k(\alpha,\beta)=m\bar k(\alpha,\beta)$ to approximate for the overall growth rate of $L'(t)$. Thus, in the large $t$ limit, 
we can apply the continuum theory introduced by Barab\'asi and Albert to calculate for an expression of the connectivity (degree) distribution for our network model \cite{Rev}. 


First, when a new vertex is added,  $m$ old vertices are connected to this vertex with preferential attachment strategy. So we can write the growth rate of the connectivity of the $i^{th}$ vertex as
      \begin{equation}
      \label{eq:eq4}
\frac{\partial k_i}{\partial t}=m\Pi (k_i)=m\frac{k_i}{\sum_{j\in V}k_j}.
      \end{equation}
Here, $\Pi (k_i)$ is the probability for the $i^{th}$ vertex (which is active) to be attached when a new vertex is introduced. $V$ is the set of active vertices in the network. Also, for the $i^{th}$ vertex, the attachments from different m edges of a new vertex are mutually exclusive. Accordingly, we have the form of Eq. \ref{eq:eq4} for the growth rate of $k_i$. Let us draw attention to the summation term in the denominator of Eq. \ref{eq:eq4}. In the large $t$ limit, we have
      \begin{equation}
      \label{eq:eq5}
\sum_{j\in V}k_j\sim2m\bar{k}(\alpha,\beta)t.
      \end{equation}
In this large $t$ limit, Eq. \ref{eq:eq4} can be written as
      \begin{equation}
      \label{eq:eq6}
\frac{\partial k_i}{\partial t}=\frac{k_i}{2\bar{k}(\alpha,\beta)t}
      \end{equation}
with a solution of
      \begin{equation}
      \label{eq:eq7}
k_i(t)=m\left ( \frac{t}{t_i} \right )^{\beta '},
      \end{equation}
where $\beta '=1/{[2\bar{k}(\alpha,\beta)]}$. Now, with the same argument as shown in Barab\'asi and Albert's previous work \cite{Rev}, we can derive the expression for the degree distribution $P(k)$ in the large $t$ limit
      \begin{equation}
      \label{eq:eq8}
P(k)=\frac{2m^{1/\beta'}t}{m_0+t}k^{-\gamma }  
\end{equation}
where $\gamma=1/\beta'+1=2\bar{k}(\alpha,\beta)+1$.  At last, we can derive the asymptotic expression for $P(k)$ as 
      \begin{equation}
      \label{eq:eq9}
P(k)\sim 
2m^{1/\beta'}k^{-\gamma }  
\end{equation}
From Eq. \ref{eq:eq9}, we conclude that the degree distribution of the networks of our model still follows a power law, as long as $t\gg \alpha ,\beta  $. Another feature found in Eq. \ref{eq:eq9} is that, the exponent $\gamma$ is always smaller than 3. This is because $\bar{k}$ is always smaller than 1 in the networks of our model, see Fig. \ref{fig3} for illustration. In general, if we have data of $\bar{k}(\alpha,\beta,t)$ for the heterogeneously controlled network, as is shown in Fig. \ref{fig3}(b), we can approximate for the time-independent $\bar{k}(\alpha,\beta)$  in the large $t$ limit. Given $\bar{k}(\alpha,\beta)$, we can readily derive the value of $\gamma$ for the degree distribution of the network.

\begin{figure} [h]
\includegraphics[width=1.0\columnwidth]{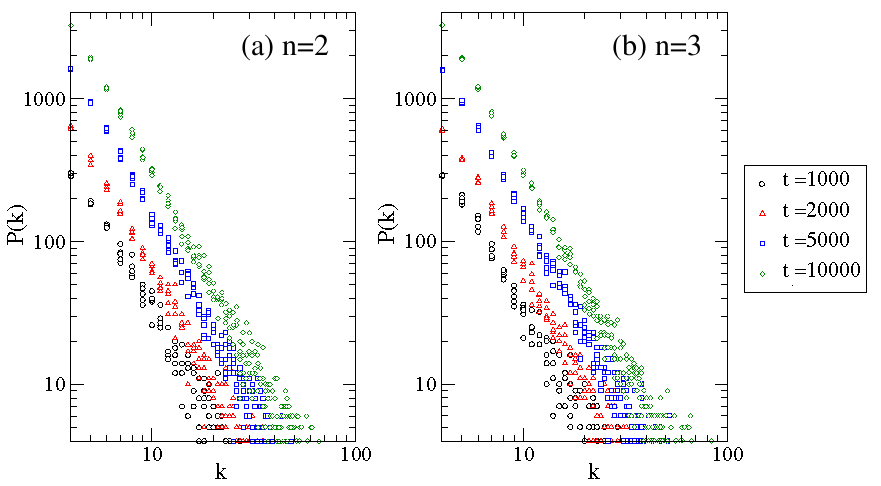}
\caption{\label{fig4} (Color online) Degree distributions of the networks with $m_0=m=4$, $\alpha=500$ and $\beta=100$. In each plot, data from five realizations of network generations are presented for a certain $t$, where $t=1000,2000,5000$ and $10000$. (a) shows the degree distributions for networks with $n=2$; while (b) shows the degree distributions for networks with $n=3$.
}
\end{figure}
Fig. \ref{fig4} shows the degree distribution for networks with fixed $\alpha$ (=500), $\beta$ (=100) and different scales of $t$. From both cases of n=2 and n=3, we find the degree distributions of our networks are kept power-law-like, even with small ratios of $t/\alpha$ and $t/\beta$ (, see the data points labeled by black circles). Also, it is found in both Fig. \ref{fig4}(a) and Fig. \ref{fig4} (b) that the slopes of $log_{10}P(k)$ and $log_{10}(k)$ are almost the same with varied $t$. At last, a comparison of Fig. \ref{fig4}(a) and Fig. \ref{fig4} (b) shows that, different $n$ values don't significantly change $P(k)$ as long as these values are kept small.

\section{Conclusions}
Real-world networks, especially the networks in human society, usually have their vertices heterogeneously controlled, which leads to repeated deactivation and reactivation of the vertices according to some standards. To understand the growth behavior of this type of networks, we proposed a model in which vertices can be repeatedly deactivated and reactivated according to its real-time status (tier) in the network. From simulations of networks with different total number of tiers, we find the total number of edges and the average geodesic length of the networks grow with different periodicities. Further, we find the growth behaviors of these two quantities resemble those of BA model, if the ratio of maximum active time $\alpha$ and maximum inactive time $\beta$ goes to infinity. This argument is confirmed by numerical simulations after the concept of normalized growth rate is introduced. Also, in the large $t$ limit, we find the degree distribution of our network is still scale free, but with an exponent $\gamma<3$. At last, we present the degree distributions in the networks with small $t$'s with data from simulations. The simulation results show scale-free-like distributions of disregard the ratio of $t/\alpha$ and $t/\beta$. More over, it is shown that the degree distribution is not significantly influenced by the total number of tiers $n$ we implement for the networks (, when $n$ is small). 


As immediate extensions of our current work, we expect to find out how the clustering coefficient of the network grows with time; we also expect to find out how the growth of the networks will be altered if we change the rule of categorization of vertices.


\begin{thebibliography}{99}

\bibitem{AB} R. Albert, A.-L. Barab‡si, Phys. Rev. Lett. {\bf 85}, 5234-5237 (2000).
\bibitem{DM} S. N. Dorogovtsev and J.F.F. Mendes, Europhys. Lett. {\bf52}, 33-39 (2000)
\bibitem{AJB} R. Albert, H. Jeong, A.-L. Barab\'asi, Nature {\bf 406} , 378Ð482 (2000). 
\bibitem{MGN} C. Moore, G. Ghoshal, M. E. J. Newman, Phys. Rev. E {\bf74}, 036121 (2006).
\bibitem{BMRS} H. Bauke, C. Moore, J.-B. Rouquier, D, Sherrington, Eur. Phys. J. B {\bf83} 519-524, (2011).
\bibitem{LH} A. Lombardi and M. Hornquist, Phys. Rev. E {\bf 75}, 56110 (2007).
\bibitem{KM} D.-H. Kim and A.E. Motter, New J. Phys. {\bf 11}, 113047 (2009). 
\bibitem{NV} T. Nepusz and T. Vicsek, Nat. Phys. {\bf 8}, 568Ð573 (2012)
\bibitem{SM} J. Sun and A. E. Motter, Phys. Rev. Lett. {\bf110}, 208701 (2013).
\bibitem{PPR} A. L. Pastore Y Piontti, C. E. La Rocca, Z. Toroczkai, L. A. Braunstein, P. A. Macri, E. Lopez, New J. Phys. {\bf10}, 093007 (2008).
\bibitem{CKM} S. P. Cornelius, W. L. Kath, and A. E. Motter, Nat. Commun. {\bf4}, 1942 (2013).
\bibitem{BAmodel} A.-L. Barab\'asi and R. Albert, Science {\bf 286}, 509--512 (1999).
\bibitem{Rev} R. Albert, A.-L. Barab‡si, Rev. Mod. Phys. {\bf74}, No. 1(2002).
\bibitem{Newman} M. E. J. Newman, J. Stat. Phys. {\bf101}, 819 (2000)
\bibitem{DA} E. W. Dijkstra, Numerische Mathematik {\bf1} 269-271 (1959)
\end{thebibliography}
\end{document}